\documentclass[preprint,nofootinbib,12pt,floatfix,superscriptaddress,aps,prd]{revtex4}
\usepackage{amsmath}
\usepackage{amssymb}
\usepackage{amsbsy}
\usepackage{amsfonts}
\usepackage{amsopn}
\usepackage{amstext}
\usepackage{graphicx}
\usepackage{amssymb}
\usepackage{amsfonts}
\usepackage{amsmath}
\usepackage{graphicx}
\usepackage{color}
\usepackage{xcolor}
\usepackage{slashed}
\usepackage{esint}
\usepackage[dvips]{epsfig}
\usepackage[dvips]{graphicx}
\usepackage{float}
\usepackage{units}
\usepackage{textcomp}
\newcommand{\beq}{\begin{equation}}
\newcommand{\eeq}{\end{equation}}
\newcommand{\bea}{\begin{eqnarray}}
\newcommand{\eea}{\end{eqnarray}}

\begin{document}

\title{Black strings from dark matter}

\author{M. S. Cunha}
\affiliation{Centro de Ci\^{e}ncias e Tecnologia, Universidade Estadual do Cear\'{a}, 60714-903, Fortaleza, Cear\'{a}, Brazil}
\email{marcony.cunha@uece.br}

\author{G. Alencar}\affiliation{Departamento de F\'isica, Universidade Federal do Cear\'a, Caixa Postal 6030, Campus do Pici, 60455-760 Fortaleza, Cear\'a, Brazil}

\author{C. R. Muniz \footnote{Corresponding author: celio.muniz@uece.br}}\affiliation{Universidade Estadual do Cear\'a, Faculdade de Educa\c c\~ao, Ci\^encias e Letras de Iguatu, 63500-000, Iguatu, CE, Brazil.}
\email{celio.muniz@uece.br}

\author{V. B. Bezerra}
\affiliation{Departamento de F\'{i}sica, Universidade Federal da Para\'{i}ba-UFPB, Caixa Postal 5008, 58051-970, Jo\~{a}o Pessoa, Para\'{i}ba, Brazil}
\email{valdir@fisica.ufpb.br}

\author{H. S. Vieira}
\affiliation{Department of Physics, Institute of Natural Sciences, Federal University of Lavras, 37200-000 Lavras, Brazil}
\affiliation{Theoretical Astrophysics, Institute for Astronomy and Astrophysics, University of T\"{u}bingen, 72076 T\"{u}bingen, Germany}
\affiliation{S\~{a}o Carlos Institute of Physics, University of S\~{a}o Paulo, 13560-970 S\~{a}o Carlos, Brazil}
\email{horacio.santana.vieira@hotmail.com}
\email{horacio_vieira@ufla.br}

\begin{abstract}

In this paper, we obtain two static black string solutions by considering as sources axisymmetric dark matter (DM) distributions in 3+1 dimensions, in two different scenarios, namely, purely isotropic and radial orbits, which are regulated by two DM parameters. Both solutions tend asymptotically to the usual static and uncharged black string vacuum solution predicted by General Relativity (GR), and each of them presents an event horizon, which is larger than the one corresponding to the vacuum solution, as it is shown. We also show that the presence of DM turns these solutions singular at their symmetry axis. Then, we calculate the Hawking temperatures of these black string horizons and discuss some of their consequences. Unlike what occurs with the static black string vacuum solution, we find that there exists a linear density of mass (or tension) remnant associated with a vanishing Hawking temperature for the solutions under consideration. Thus, we analyze how the presence of DM affects the occurrence of the remnants. Further, we calculate other thermodynamic quantities, namely entropy, heat capacity, and free energy per unit length, showing that thermal phase transitions can occur in the presence of DM. We also analyze the weak (and null) energy conditions and conclude that DM does not behave like an exotic fluid. Finally, the corresponding stationary solutions are obtained, as well as their new tensions as functions of both the mass and angular momentum of the black strings. 
\vspace{0.75cm}
\end{abstract}

\pacs{72.80.Le, 72.15.Nj, 11.30.Rd}

\maketitle
\section{Introduction}
Besides spherical symmetry, the cylindrical one has played an important role in GR since its early years. As examples of this,  we can mention the static solutions obtained by Levi-Civita and Weyl \cite{Levi,Weyl}, Chazy and Curzon \cite{Chazy,Curzon}, as well as their rotating counterpart found by Lewis \cite{Lewis}. Among other solutions with cylindrical symmetry, we can mention the one corresponding to the spacetime of a cosmic string \cite{Vilenkin,Gott} as well as those corresponding to G\"{o}del and Krasinsky spacetimes \cite{Godel,Krasinski}.

Black strings are more recent symmetrically cylindric solutions obtained from Einstein's equations in a D-dimensional spacetime \cite{duff1988}. These vacuum objects are generalizations of black holes with translation symmetry where the event horizon is topologically equivalent to $S_2 \times R$ (or $S_2 \times S_1$ in the case of black rings) and the spacetime is asymptotically $M_{D-1} \times S_1$ for a zero cosmological constant or $AdS_{D-1} \times S_1$ for a negative one \cite{emparanPRL,ijmpa2011}. In despite of the hoop conjecture formulated by Thorne \cite{Thorne} in which the gravitational collapse of a massive star will produce a black hole only if the mass $M$ is smashed within a region with circumference $C<4\pi M$ in all directions, a counterexample of a 4-D vacuum black hole solution of the Einstein field equations with cylindrical symmetry was given in \cite{Lemos1,Lemos2}. Using the Hamiltonian formulation of GR, it was managed to define mass and angular momentum for the obtained solution. Since then such an object has been investigated in several scenarios, as in massive gravity \cite{Ghosh,Hendi}, noncommutative geometry \cite{Singh}, dynamical Chern-Simons gravity \cite{Pino}, mimetic gravity \cite{Ahmad}, with anisotropic quintessence fluid \cite{Sabir}, and non-local gravity \cite{Celio}, among others.

The standard model of cosmology, supported by observational data, states that DM constitutes about 29.6\% of the total content of energy-matter in the Universe \cite{Planck:2018vyg}. Galaxies and their clusters present within and around them large quantities of DM, which strongly contributes to the formation, evolution, and coalescence of such structures using the gravitational interaction \cite{Trujillo-Gomez:2010jbn}, and even the ``cosmic web'', the filamentous structure observed on a large scale with galaxy clusters occupying its intersections, occurs under the presence of DM \cite{Kim,Callum}.

Based on the points raised above, there is enough motivation to consider that black strings sourced by axisymmetric dark matter distribution should have done, if they existed, an interesting role concerning structure formation at cosmological scale \cite{Eisenstein}. For this reason, such objects must be investigated. Besides this, we take also into account the recently raised possibility of the occurrence of a negative cosmological constant in the dark sector of the Universe \cite{Rodrigo}, a fundamental ingredient of black strings. Then, we will study several features involving such objects and the influence of DM on them, such as the presence of horizons, singularities, the Hawking temperature, the existence of remnants, as well as the change in their thermodynamic properties, including the thermal (un)stability and occurrence of phase transitions, as compared to those associated to the vacuum solution. We will also discuss the energy conditions related to DM matter supporting static black strings. For the sake of completeness, we will obtain and discuss the stationary counterparts of the static solutions, by adding angular momentum to the static spacetime, using an appropriate coordinate transformation \cite{Lemos2}.

The paper is organized as follows: In section II we find the static black string solutions sourced by axisymmetric dark matter and investigate their horizons, thermodynamics, and energy conditions. In section III we present and discuss briefly the corresponding stationary solutions. Finally, in section IV we offer the conclusions about the results obtained.

\section{Black string solutions in the presence of dark matter}
We start with Einstein's equations of GR in order to find the geometry of the black string sourced by dark matter, with the cosmological constant term, namely
\begin{equation}\label{EME}
	G^{\mu}_{\nu}+g^{\mu}_{\nu}\Lambda=\kappa T^{\mu}_{\nu},
\end{equation}
Considering the cylindrical symmetry around the $z$ axis, we get that the dark matter density profile, according to an isothermal distribution of radial velocities, is given by \cite{Eisenstein}
\begin{equation}\label{Rho}
\rho(r)=\rho_s\frac{(2-\beta)^2\left(\frac{r}{R_s}\right)^{-\beta}}{\left[1+\left(\frac{r}{R_s}\right)^{2-\beta}\right]^2},
\end{equation}
where $0\leq\beta\leq 1$, with $\beta=0$ ($\beta=1$) for purely isotropic (radial) orbits, with $R_s$ ($\rho_s$) being an arbitrary scale factor (density) related to each specific scenario. Henceforth, we will adopt the relation $\Lambda=-3/\ell^2$, where $\ell$ is a fundamental length of the model.
Let us assume that the black string metric has the following form
\begin{equation}\label{BlackStringMetric}
ds^2=-f(r)dt^2+\frac{dr^2}{f(r)}+ r^2d\phi^2+ \frac{r^2}{\ell^2} dz^2.
\end{equation}
Einstein's tensor is then given by
\bea
G^{\mu}_{\nu}=
\left(
\begin{array}{cccc}
 \frac{r f'(r)+f(r)}{r^2} & 0 & 0 & 0 \\
 0 & \frac{r f'(r)+f(r)}{r^2} & 0 & 0 \\
 0 & 0 & \frac{r f''(r)+2 f'(r)}{2 r} & 0 \\
 0 & 0 & 0 & \frac{r f''(r)+2 f'(r)}{2 r} \\
\end{array}
\right)
\eea

Solving the $ 00$ component of Einstein's equation, we find the following results
\begin{equation}\label{metrics}
f(r)=\left\{ \begin{array}{cc}
-\frac{4\mu\ell}{r}+\frac{r^2}{\ell^2}+\frac{2\kappa  \text{$\rho_s$} R_s^2}{ 1+(r/R_s)^2}-\frac{2\kappa  \text{$\rho_s$} R_s^3 \tan ^{-1}\left(\frac{r}{R_s}\right)}{ r}, & \ \ \beta=0;\\
-\frac{4\mu\ell}{r}+\frac{r^2}{\ell^2}-\frac{\kappa  \rho_s R_s^3}{r(1+r/R_s)}- \frac{\kappa\rho_s R_s^3 \log (1+\frac{r}{R_s})}{r}, & \ \ \beta=1,
\end{array}\right.
\end{equation}
where the integration constants were chosen in order to appear the linear density of the black string, $\mu$, so that in absence of dark matter, $\rho_s=0$, we retrieve Lemos' vacuum solution \cite{Lemos1,Lemos2}. For $r$ much greater than the arbitrary scale factor $R_s$ ($r/R_s>>1$), the approximate solution up to third order in it and taking $\beta = 1$ is given by
\begin{equation}
f(r)\approx -\frac{4\mu\ell}{r}+\frac{r^2}{\ell^2}-\frac{\kappa\rho_sR_s^3}{r}\log{\left(\frac{r}{R_s}\right)}.
\end{equation}
On the other hand, for the solution associated with $\beta=0$, the corresponding expansion keeps the same structure of the black string in the vacuum, namely,
\begin{equation}
f(r)\approx -\frac{4\tilde{\mu}\ell}{r}+\frac{r^2}{\ell^2},
\end{equation}
with $\tilde{\mu}=\mu+\frac{\pi\kappa \rho_sR_s^3}{4\ell}.$

\begin{figure}[h!]
    \centering
            \includegraphics[width=0.50\textwidth]{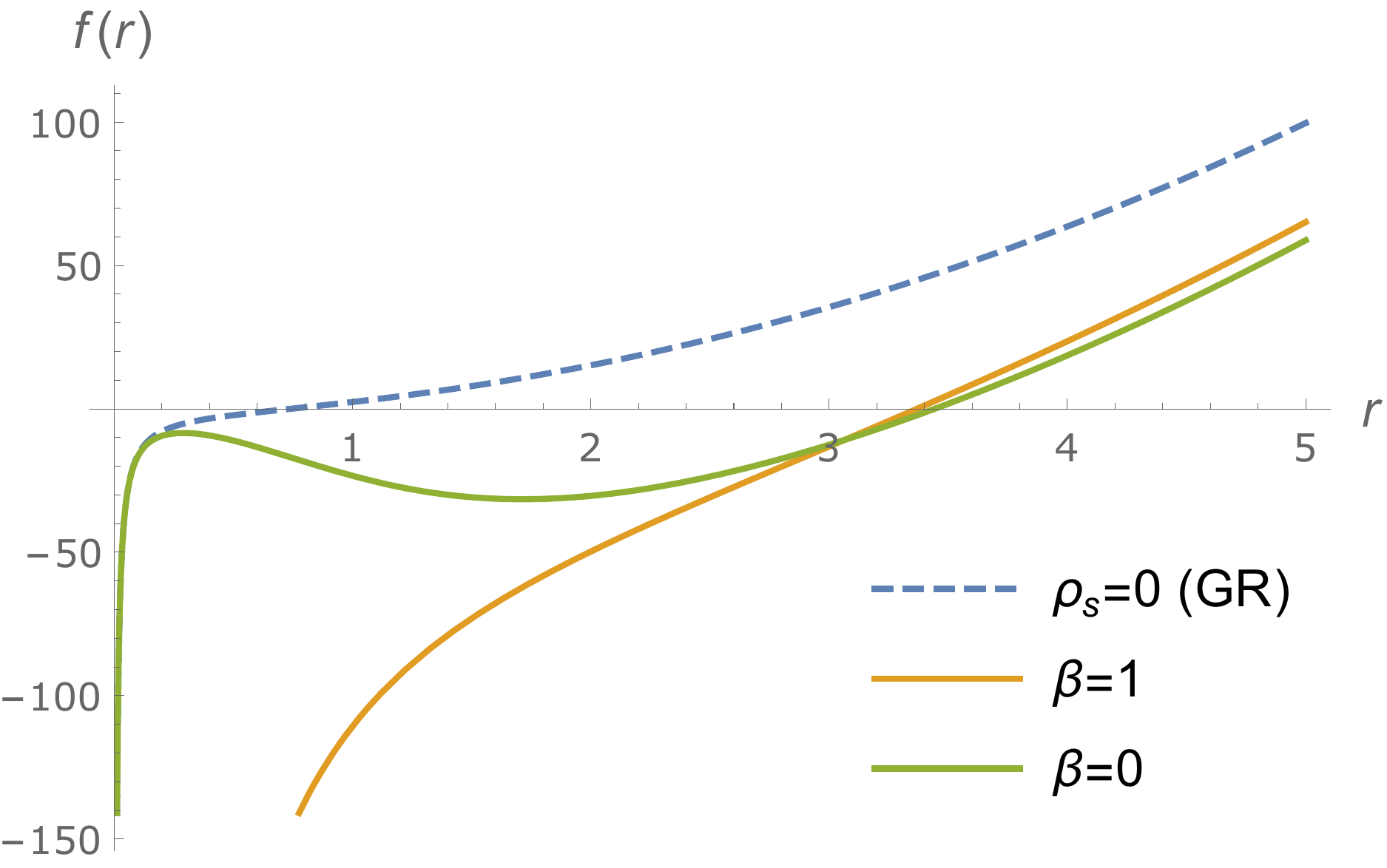}
        \caption{The metric function in terms of the radial coordinate for two different values of $\beta$ ($\rho_s=0$ is the vacuum solution of GR). The parameter settings are $\ell=0.5$, $\mu=0.8$, $\rho_{s}=1.2$ and $R_{s}=1.5$, in Planck units, with $\kappa=8\pi$.}
    \label{BSDMFig1}
\end{figure}

\begin{figure}
\centering
           \includegraphics[width=0.55\textwidth]{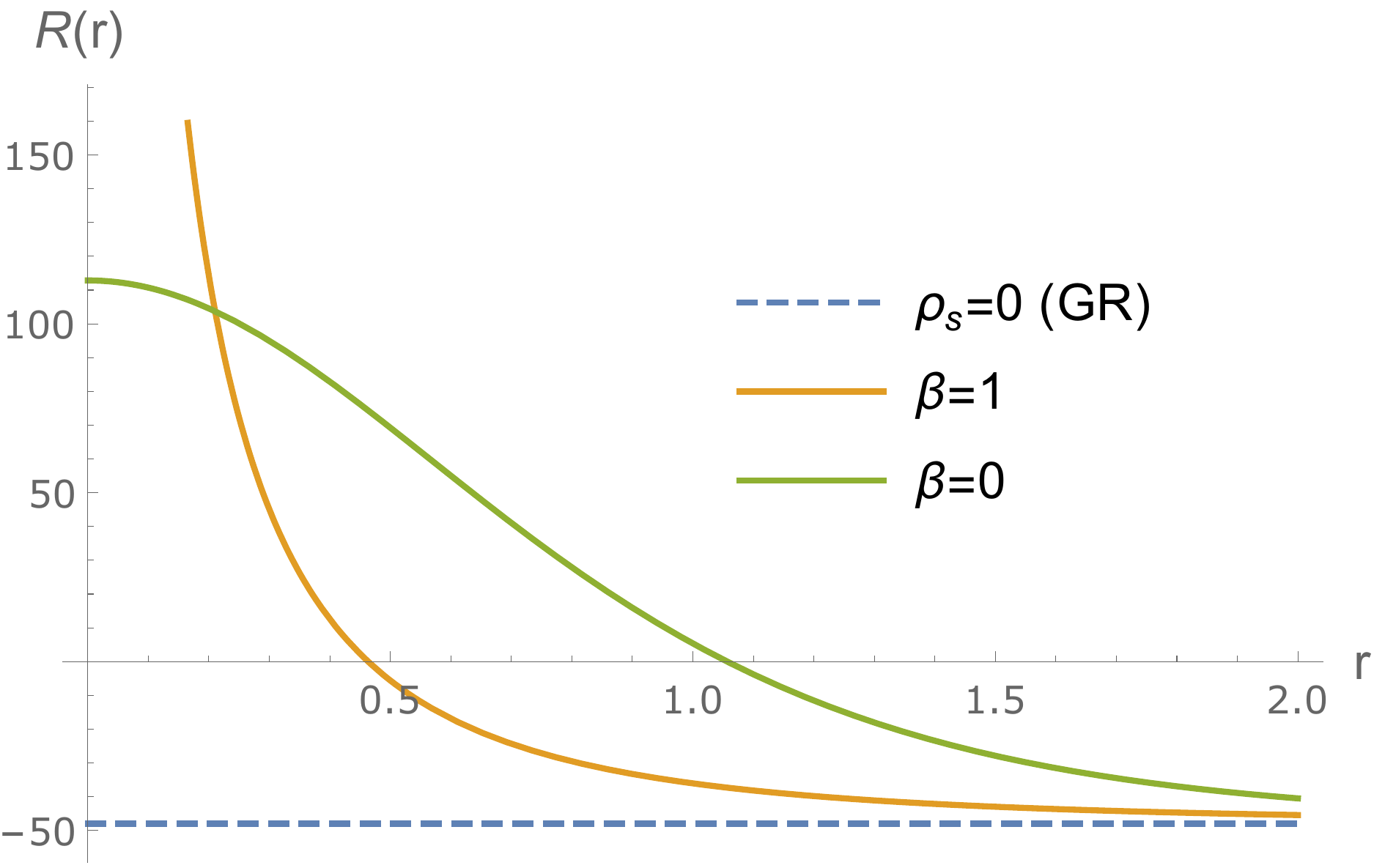}
       \caption{Scalar of curvature. The parameters settings are $\ell=0.5$, $R_{s}=0.5$, $\mu=0.8$ and $\rho_s=0.4$ in Planck units, with $\kappa=8\pi$.}
   \label{ScalCDMFig1}
\end{figure}

In Fig.\ref{BSDMFig1}, we depict the metric function $f(r)$ for some values of $\beta$. Notice the existence of only one event horizon for both the solutions, namely, $\beta = 0$ and $\beta = 1$, which is larger than the one corresponding to the vacuum GR solution ($\rho_s=0$). We also can show that the greater the concentration of DM, the greater these horizon radii are.
It is worth studying how DM influences the spacetime singularity. The expressions for the Ricci scalar are
\bea
R_{\beta=0} &=& \frac{4R_s^6\left(4\kappa l^2 \rho_s-3\right)-36R_s^4 r^2 -36R_s^2 r^4-12r^6}{l^2 \left(r^2+R_s^2\right){}^3} \\
R_{\beta=1} &=& \frac{3 \kappa  l^2 R_s^4 \rho_s+R_s^3 \left(\kappa  l^2 \rho _s-12\right)r-36R_s^2 r^2-36R_s r^3-12r^4}{l^2r\left(r+R_s\right)^3}
\eea\\
We can see that, for $\rho_s=R_s=0$, we recover the constant value of our vacuum solution. The plots are given in Fig \ref{ScalCDMFig1}.
In other to study the singularity, we expand the above expressions for small values of $r$. We have, respectively
\begin{eqnarray}
R_{\beta=0}&=& \frac{4(-3+4\ell^2\kappa\rho_s)}{\ell^2}-\frac{48 \kappa \rho_s} {R_s^2 }\,r^2+\mathcal{O}(r)^4,
\end{eqnarray}

\begin{eqnarray}
R_{\beta=1}&=& \frac{3 \kappa \rho_s R_s} {r} -\frac{12+8\ell^2\kappa\rho_s}{\ell^2}+\frac{15 \kappa \rho_s}{R_s}\,r+\mathcal{O}(r)^3.
\end{eqnarray}

  Therefore, in the presence of DM with $\beta=1$, the Ricci scalar becomes singular. {Can we say that DM strengthens the singularity? This can be confirmed from other scalars. Below we give the the Ricci squared $R^{(2)}=R_{\mu\nu}R^{\mu\nu}$, and  Kretshmann scalar $K=R_{\mu\nu\alpha\beta}R^{\mu\nu\alpha\beta}$. 
\\
\\
For the squared Ricci scalar, we have
\bea
R^{(2)}_{\beta=0}= \frac{4}{l^4 \left(r^2+R^2\right)^6}  \Big\lbrace 9r^{12}+54 r^{10} R^2 + 135 r^8 R^4 +12 r^6 R^6 \left(15-2l^2 \kappa\rho \right) \hspace{3cm}\nonumber\\
+r^4 R^8 \left(16 l^4 \kappa^2\rho^2 - 72 l^2\kappa \rho +135\right)+ 18 r^2 R^{10} \left(3-4l^2 \kappa \rho \right)+R^{12} \left(3-4l^2 \kappa \rho \right)^2 \Big\rbrace \hspace{1cm}
\eea
and
\bea
R^{(2)}_{\beta=1} = \frac{1}{2 l^4 r^2 (r+R)^6}\Big\lbrace 72r^8+432 Rr^7+1080 R^2 r^6-12R_s^3 \left(\kappa  l^2\rho-120\right)r^5 \hspace{2.69cm}\nonumber\\
-72 R_s^4  \left(\kappa  l^2 \rho -15\right)r^4-144 R_s^5 \left(\kappa  l^2 \rho -3\right)r^3 + R^6 \left(5 \kappa ^2 l^4 \rho ^2-120 \kappa  l^2 \rho +72\right)r^2\hspace{1.1cm} \nonumber\\
+6 \kappa  l^2 \rho_s R^7 \left(\kappa  l^2 \rho -6\right)r + \kappa^2 l^4\rho_s^2 R_s^8 \Big\rbrace 
\hspace{8.4cm}
\eea
 As expected, for $\rho_s=R_s=0$, we recover a constant value. The plots are given in Fig. \ref{Ricci_sq}.  Expanding the above expressions, for small values of $r$, we have, respectively 
 \bea
R^{(2)}_{\beta=0}&=&\frac{4 \left(3-4 \kappa  l^2 \rho _s\right){}^2}{l^4}-\frac{96 \left(4\kappa^2 l^2\rho_s^2-3 \kappa  \rho _s\right)}{l^2 R_s^2}\,r^2 + \mathcal{O}(r)^3\\
\nonumber\\
R^{(2)}_{\beta=1} &=& \frac{5 \kappa ^2 R_s^2 \rho _s^2}{2 r^2}-\frac{12 \kappa ^2 R_s \rho _s^2+18 \kappa  R_s \rho _s}{l^2 r}+\frac{36 \kappa ^2 l^4 \rho _s^2+48 \kappa  l^2 \rho _s+36}{l^4}+ \mathcal{O}(r)
\eea
Therefore, for $\beta=1$, DM strengthens the singularity of the Ricci squared. 

 Finally, we will analyze the Kretshmann scalar. It is given by
\begin{eqnarray}
K_{\beta=0}=\frac{8}{l^4 r^6 \left(r^2+R_s^2\right)^6} \Big\lbrace  3r^{18}+18 R_s^2 r^{16}+45 R_s^4 r^{14}+\left[\mu^2 l^6 + R_s^6 \left(60-8 l^2 \kappa\rho_s \right)\right]r^{12}\hspace{5.2cm}\nonumber\\
-72l^5\mu\kappa\rho_s R_s^4 r^{11} + \left[144 l^6 \mu^2 R_s^2+R_s^8\left(70 \kappa ^2 l^4 \rho_s ^2-24 \kappa  l^2 \rho_s +45\right)\right]r^{10}-280l^5\mu\kappa\rho_s R_s^6 r^9 \hspace{4cm}\nonumber\\
+6 R_s^4 \left[60\mu^2 l^6+R_s^6 \left(16 \kappa ^2 l^4 \rho_s^2-4 \kappa  l^2 \rho_s +3\right)\right]r^8 - 432l^5\mu\kappa\rho_s R_s^8r^7\hspace{7.2
cm}\nonumber\\
 +R_s^6 \left[480\mu^2 l^6+R_s^6 \left(84\kappa^2 l^4 \rho_s^2-8 \kappa  l^2 \rho_s +3\right)\right]r^6 - 336 l^5\mu\kappa\rho_s R_s^{10}r^5\hspace{7.2cm}\nonumber\\
+8 l^4 R_s^8\left(45 l^2 \mu ^2+4 \kappa ^2 \rho_s ^2 R_s^6\right)r^4-136 l^5\mu\kappa\rho_s R_s^{12}r^3 + 6 l^4 R_s^{10} \left(24 l^2\mu^2+\kappa^2\rho_s^2 R_s^6\right)r^2\hspace{4.6cm}\nonumber\\
- 24 l^5 \mu\kappa\rho_s R_s^{14} r+24l^6\mu^2 R_s^{12}+ 6l^4\kappa^2\rho_s^2 R_s^6 \left(r^2+R_s^2\right)^6 \tan^{-1}\left(\frac{r}{R_s}\right)^2\hspace{7.2cm}\nonumber\\
+4 l^4 \kappa\rho_s R_s^3\tan^{-1}\left(\frac{r}{R_s}\right) \left[6 l \mu  \left(r^2+R_s^2\right)^6-\kappa  \rho_s  r R_s^4 \left(r^2+R_s^2\right)^3 \left(9 r^4+8 R_s^2 r^2 + 3 R_s^4\right)\right]\Big\rbrace, \hspace{3.8cm}\nonumber\\
\nonumber\\
K_{\beta=1}=\frac{1}{{l^4 r^6 (r\!+\!R_s)^6}} \Big\lbrace 24r^{12}\!+\!144 R_sr^{11} \!+\! 360R_s^2r^{10}\!-\!4R_s^3\left(\kappa l^2\rho_s\!-\!120\right)r^9\!-\!24R_s^4\left(\kappa l^2\rho_s \!-\! 15\right)r^8 \hspace{3cm}\nonumber\\
-48R_s^5 \left(\kappa l^2\rho_s-3 \right)r^7 + \left[192\mu^2 l^6-80\kappa\mu l^5\rho_s R_s^3+R_s^6 \left(13 \kappa ^2 l^4 \rho_s^2-40\kappa l^2 \rho_s +24\right)\right]r^6 \hspace{4.6cm}\nonumber\\
+6 \left(192~l^6 \mu^2 R_s-32 \kappa  l^5 \mu\rho_s R_s^4-l^4\kappa^2 \rho_s^2 R_s^7-2 \kappa l^2\rho_s R_s^7\right)r^5\hspace{8.8cm}\nonumber\\
+l^4 R_s^2\left(2880~l^2 \mu^2\!+\!96 \kappa l\mu\rho_s  R_s^3\!-\!35 \kappa^2\rho_s^2 R_s^6\right)r^4\!+\!4l^4 R_s^3 \left(960~l^2 \mu ^2\!+\!184 \kappa l\mu  \rho_s R_s^3\!+\!\kappa^2\rho_s^2 R_s^6\right)r^3
\hspace{3.4cm}\nonumber\\
+12l^4 R_s^4 \left(240~l^2\mu^2\!+\!76\kappa l\mu\rho_s R_s^3+5 \kappa^2 \rho_s^2 R_s^6\right)r^2
+48 l^4 R_s^5 \left(24 l^2 \mu^2+10 \kappa l\mu\rho_s R_s^3+\kappa^2\rho_s^2 R_s^6\right)r
\hspace{3.2cm}\nonumber\\
+4 \kappa  l^4 \rho_s R_s^3 \log \left(\frac{r\!+\!R_s}{R_s}\right) \left[24 l \mu  (r\!+\!R_s)^6+\kappa  \rho_s  R_s^3 \left(-5 r^3+3 r^2 R_s+12 r R_s^2+6 R_s^3\right) (r+R_s)^3\right]\hspace{3cm}\nonumber\\
+12 \kappa ^2 l^4 \rho_s ^2 R_s^6 (r+R_s)^6 \log ^2\left(\frac{r+R_s}{R_s}\right)+12~l^4 R_s^6 \left(4 l \mu +\kappa  \rho  R_s^3\right)^2\Big\rbrace\hspace{7.5cm}\nonumber1
\eea
Again, for $\rho_s=R_s=0$,  we get the usual $1/r^6$ singularity. The plots are given in Fig. \ref{Kretsh}. Kretshmann's scalars approximations for $r \ll 1$ are given by
\bea
K_{\beta=0}&=&\frac{192\ell^2\mu^2}{r^6}-\frac{768\,\kappa\ell\mu\rho _s}{5R_s^2}\frac{1}{r}+\frac{72-192\ell^2\kappa\rho_s+128\ell^4\kappa^2\rho_s^2}{3\ell^4}+\frac{3840\,\kappa\ell \mu \rho_s}{7R_s^4}\,r+\mathcal{O}(r)^2\hspace{1.8cm} \\
\nonumber\\
K_{\beta=1}&=&\frac{[\mu\kappa\rho_s\,l R_s]_{-6}}{r^6}+\frac{[\mu\kappa\rho_s\,l R_s]_{-2}}{r^2}+\frac{[\mu\kappa\rho_s\,l R_s]_{-1}}{r}+[\mu\kappa\rho_s\,l R_s]+[\mu\kappa\rho_s\,l R_s]_1 r+\mathcal{O}(r)^2
\eea
where $[\mu\kappa\rho_s\,l R_s]$ are combinations of these parameters respectively to the order expansion. Therefore, and curiously, DM does not strengthen the singularity of the Kretshmann scalars. They keep the behavior $1/r^6$.


\begin{figure}
\centering
\begin{minipage}{0.49\linewidth}
        \centering
           \includegraphics[width=1.\textwidth]{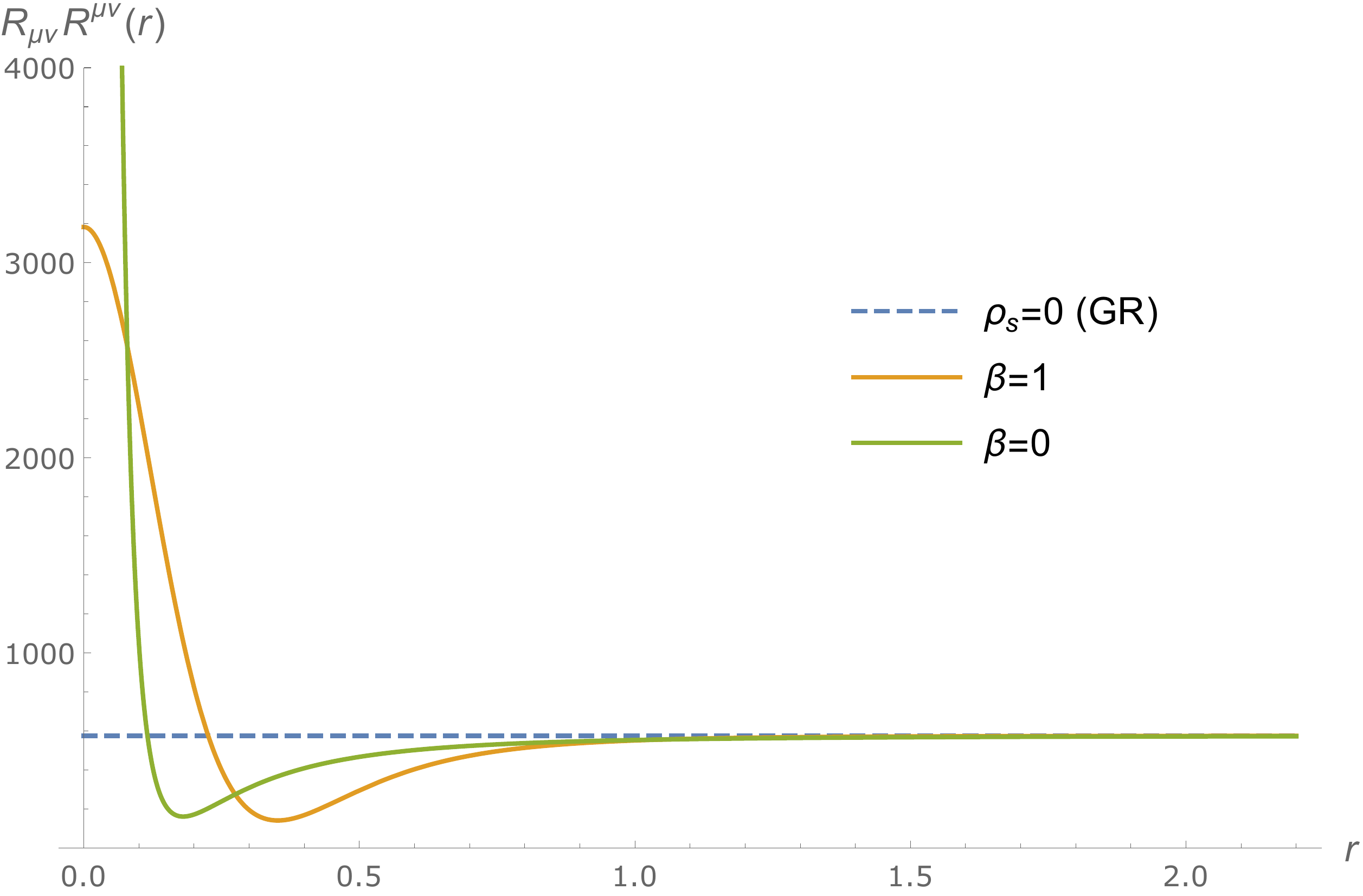}
   \label{Ricci_sq}
\end{minipage}
\begin{minipage}{0.49\linewidth}
        \centering
       \includegraphics[width=1.\textwidth]{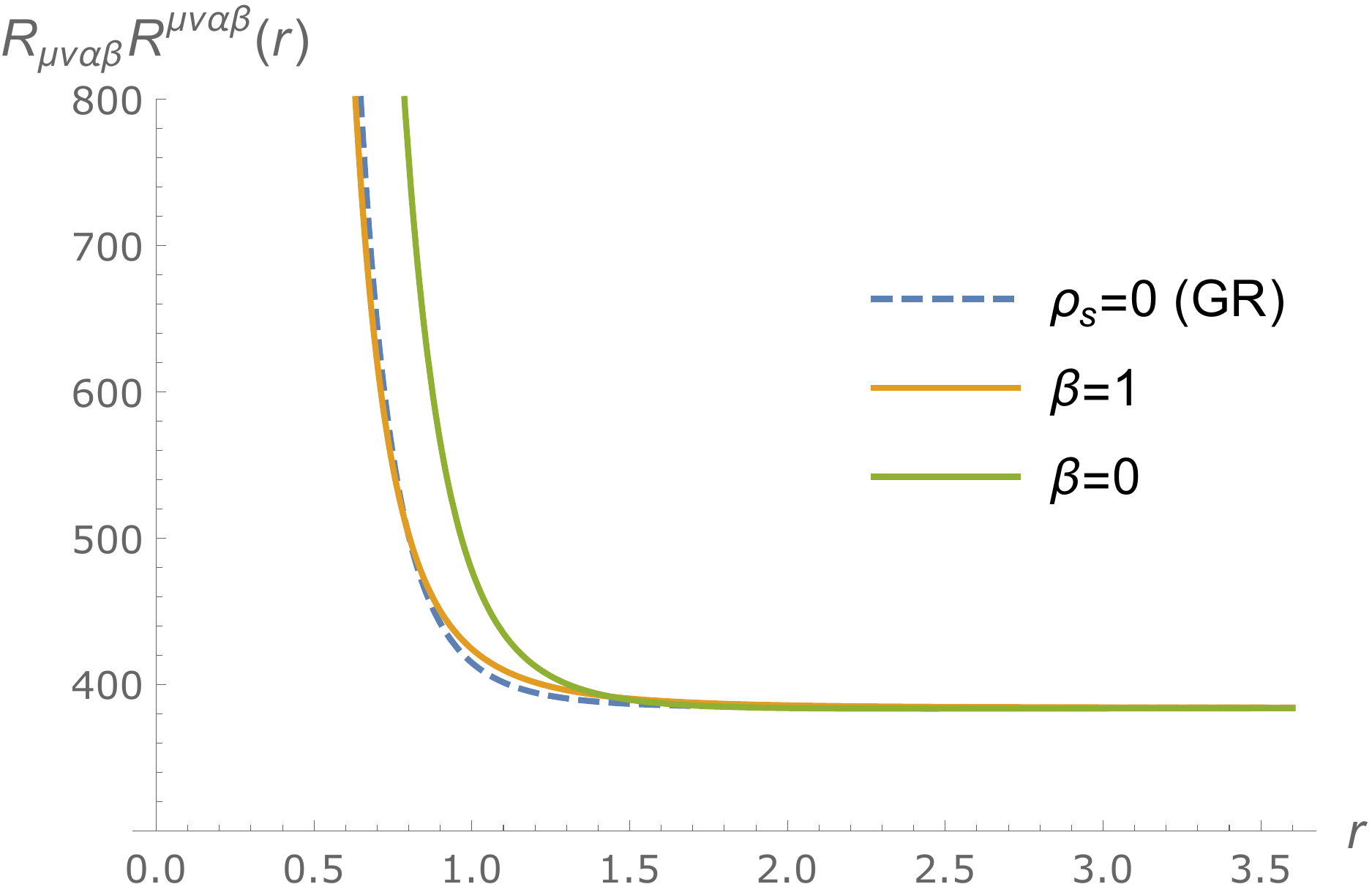}
         \label{Kretsh}
\end{minipage}
        \caption{Squared Ricci scalar (left) and Kretshmann scalar (right). The parameters settings are $\ell=0.5$, $R_{s}=0.5$, $\mu=0.8$ and $\rho_s=0.4$ in Planck units, with $\kappa=8\pi$.}
\end{figure}
The Hawking temperature associated with the obtained black string solutions can be computed employing $T_H=\frac{1}{4\pi}f'(r_h)$, where $r_h$ is the position of the event horizon, calculated from $f(r)=0$. Then we find, for $\beta=0$,
\begin{equation}
T_H = \frac{r_h\left[6 r_h^2 R_s^2+3 r_h^4+R_s^4 \left(3-4 \kappa  \ell^2 \rho _s\right)\right]}{4 \pi  \ell^2 \left(r_h^2+R_s^2\right){}^{\!2}},
\end{equation}
and for $\beta=1$,
\begin{equation}
T_H=\frac{6 r_h^2 R_s+3 r_h R_s^2+3 r_h^3-\kappa  \ell^2 R_s^3 \rho _s}{4 \pi\ell^2 \left(r_h+R_s\right){}^{\!2}}.
\end{equation}

 It is possible to see that in absence of DM ({\it i.e.}, when $\rho_s=R_s=0$), we obtain $T_H =3r_h/4\pi\ell^2$ for both cases, like in the vacuum static black string case.

 In Fig. \ref{BSDMFig2} we depict the Hawking temperature as a function of the event horizon radius. Notice the convergence to the Hawking temperature associated to the solution obtained by Lemos \cite{Lemos1,Lemos2}, for large radii of the event horizon when compared to the dark matter arbitrary scale factor, $R_s$.

\begin{figure}
\centering
            \includegraphics[width=0.55\textwidth]{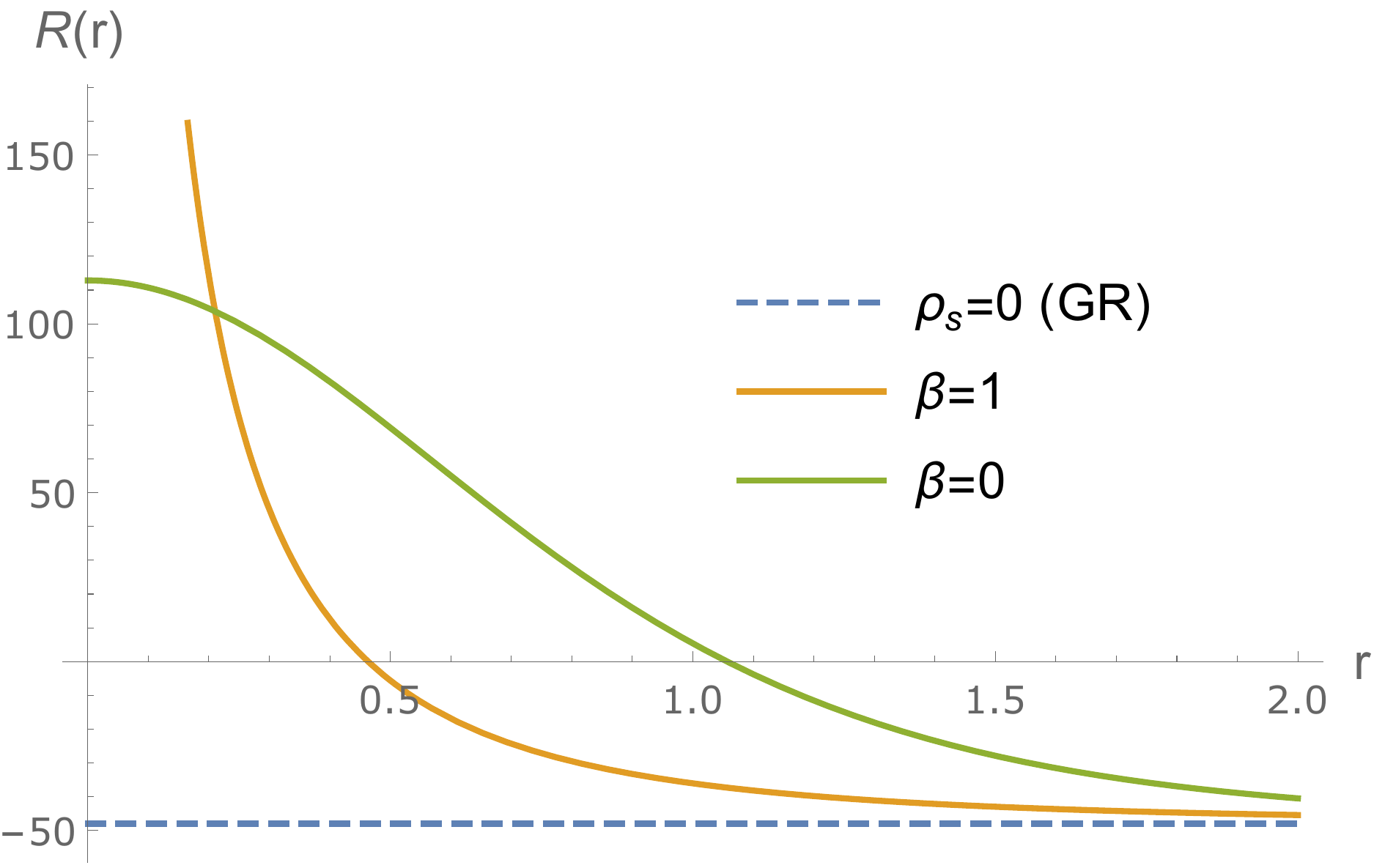}
        \caption{Hawking temperature as a function of the event horizon radius, $r_h$, for the three cases. The parameter settings are $\ell=0.5$, $\rho_{s}=1.2$, $R_{s}=1.5$, and $\kappa=8\pi$ in Planck units.}
    \label{BSDMFig2}
\end{figure}

Note that there is a critical event horizon position for which the Hawking temperature vanishes and the evaporation process halts giving rise to a remnant linear density of mass. For the black string solution in which $\beta=0$, this critical horizon radius is exactly given by
 \begin{equation}
 r^{rem}_h=R_s\sqrt{\left(\frac{4\kappa\ell^2\rho_s}{3}\right)^{\!1/2}-1}.
 \end{equation}
It is worth calling attention to the fact that for low concentrations of DM, namely, $\rho_s\leq \frac{3}{4\kappa\ell^2}$, the remnant black string would no longer exist. On the other hand, for the $\beta=1$ case, one can show that such a remnant always will exist, for any DM density, $\rho_s\neq 0$. In this case, the expression for the critical horizon radius is more involved, that is,
\bea
r_h^{rem}=\frac{1+\left(1+\frac{9}{2} \kappa\ell^2\rho_s+\frac{3}{2} \sqrt{9\kappa^2\ell^4 \rho_s^2+4\kappa\ell^2\rho _s}\right)^{2/3}}{\left(1+\frac{9}{2}\kappa\ell^2\rho_s + \frac{3}{2}\sqrt{9 \kappa^2\ell^4\rho_s^2 + 4\kappa\ell^2\rho_s}\right)^{1/3}}\frac{R_s}{3}-\frac{2R_s}{3}
\eea

and for $(\rho_s, R_s)\to 0$, we have $r^{rem}_h\approx R_s (4\kappa\ell^2\rho_s)^{1/2}/3$.

Calculating the entropy per unit length of the black string sustained by DM, we get the following result
\begin{equation}
s=\int_0^{r_h}\frac{d\mu}{dr_h^{'}}\frac{dr_h^{'}}{T_H}=\frac{\pi r_h^2}{2\ell},
\end{equation}
for both solutions, which is in accordance with the result found for the vacuum static black string.

Regarding the heat capacity per unit length, $c=(\partial \mu/\partial T_H)=[(\partial\mu/\partial r_h)(\partial T_H/\partial r_h)^{-1}]$, this is, 

\bea
c_{\beta=0} &=& \frac{\pi  \left(r^2+R_s^2\right) \left[r^2 R_s^4 \left(3-4 \kappa  \ell^2 \rho _s\right)+3 r^6+6 r^4 R_s^2\right]}{\ell \left[3 r^2 R_s^4 \left(4 \kappa  \ell^2 \rho _s+3\right)+R_s^6 \left(3-4 \kappa  \ell^2 \rho _s\right)+3 r^6+9 r^4 R_s^2\right]}\\
c_{\beta=1} &=& \frac{\pi  r \left(r+R_s\right) \left(-\kappa  \ell^2 R_s^3 \rho _s+3 r^3+6 r^2 R_s+3 r R_s^2\right)}{\ell \left[R_s^3 \left(2 \kappa  \ell^2 \rho _s+3\right)+3 r^3+9 r^2 R_s+9 r R_s^2\right]}
\eea
}%
we depict this quantity in the left panel of Fig. (\ref{figurasminipg1}) for the black string solutions with DM, and compare it with the vacuum solution, where $c=\pi r_h^2/\ell$. {Notice the phase transitions of second and zero orders in the case of $\beta=0$ solution, when the curve abruptly crosses the region with local thermal stability to the unstable one, and then smoothly tends to the stable one newly. The $\beta=1$ solution only exhibits a transition of zero order, from local thermal instability to stability}. Both the heat capacities of the found black string solutions converge to that related to the vacuum solution in GR, for large values of $r_h$, $c=\pi r_h^2/\ell$. 

On the other hand, we can conclude by analyzing the right panel of Fig. \ref{figurasminipg1} that the free energy per unit length, $F=\mu-T_H s$, reveals a zero-order phase transition for $\beta=1$ solution at $r_h\approx 1.07$, with the system going from the global thermal instability ($F>0$) to the stability ($F<0$). Notice that, in this case, the free energy of the black string tends to a constant value, $F\to \kappa R_s^3 \rho_s/(4 \ell)$, when $r_h\to 0$. Considering the $\beta=0$ solution, it exhibits a phase transition of zero order at $r_h\approx 1$, above which the black string becomes thermally stable ($F<0$). For both black string solutions sourced by dark matter, the thermodynamic quantity tends to the value of the vacuum solution, $F=-r_h^3/8\ell^3$, for large event horizons.

  \begin{figure}[!ht]
    \centering
    \begin{minipage}{0.45\linewidth}
        \centering
        \includegraphics[width=1.0\textwidth]{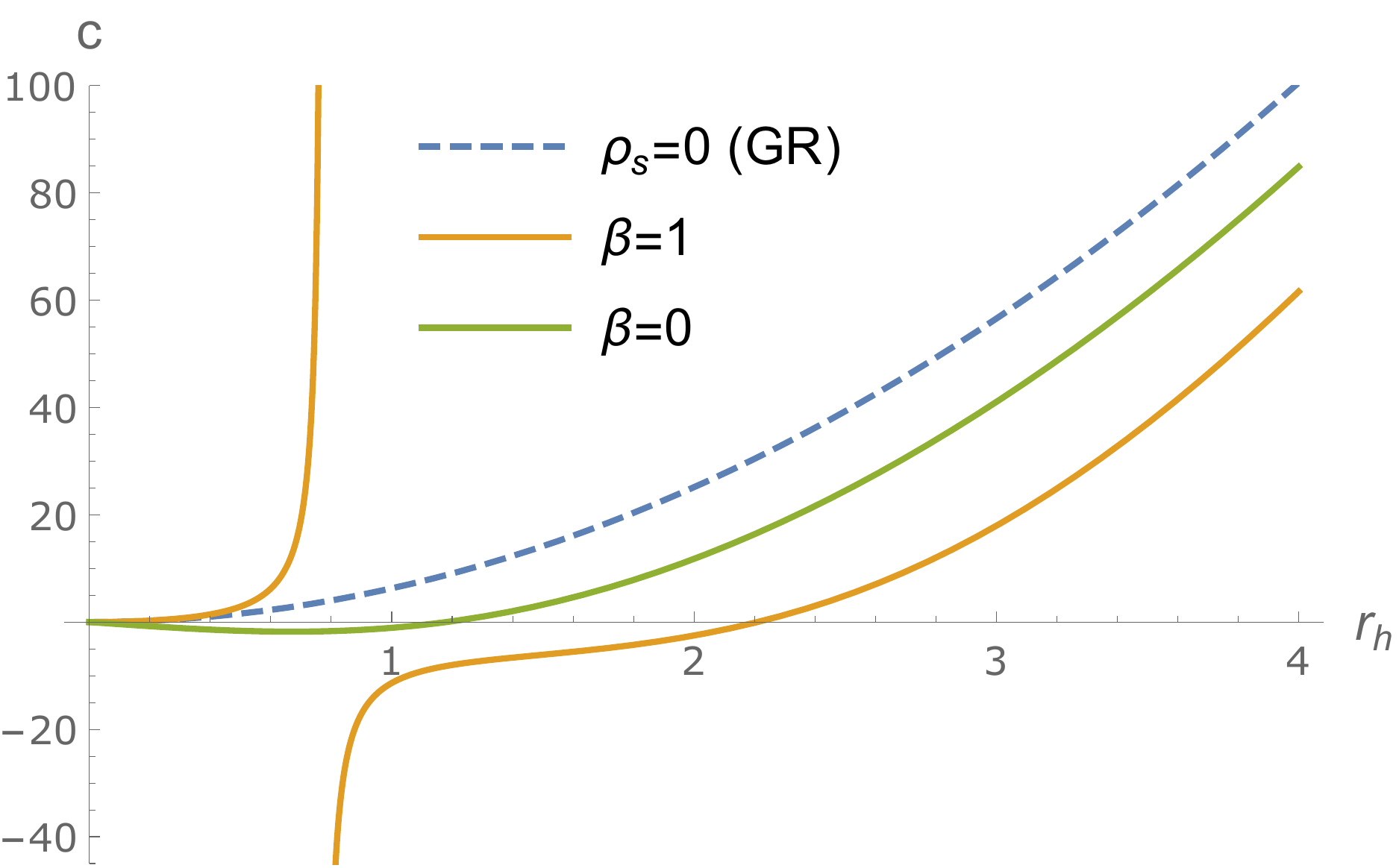}
            \end{minipage}\hfill
    \begin{minipage}{0.45\linewidth}    
        \centering
        \includegraphics[width=1.0\textwidth]{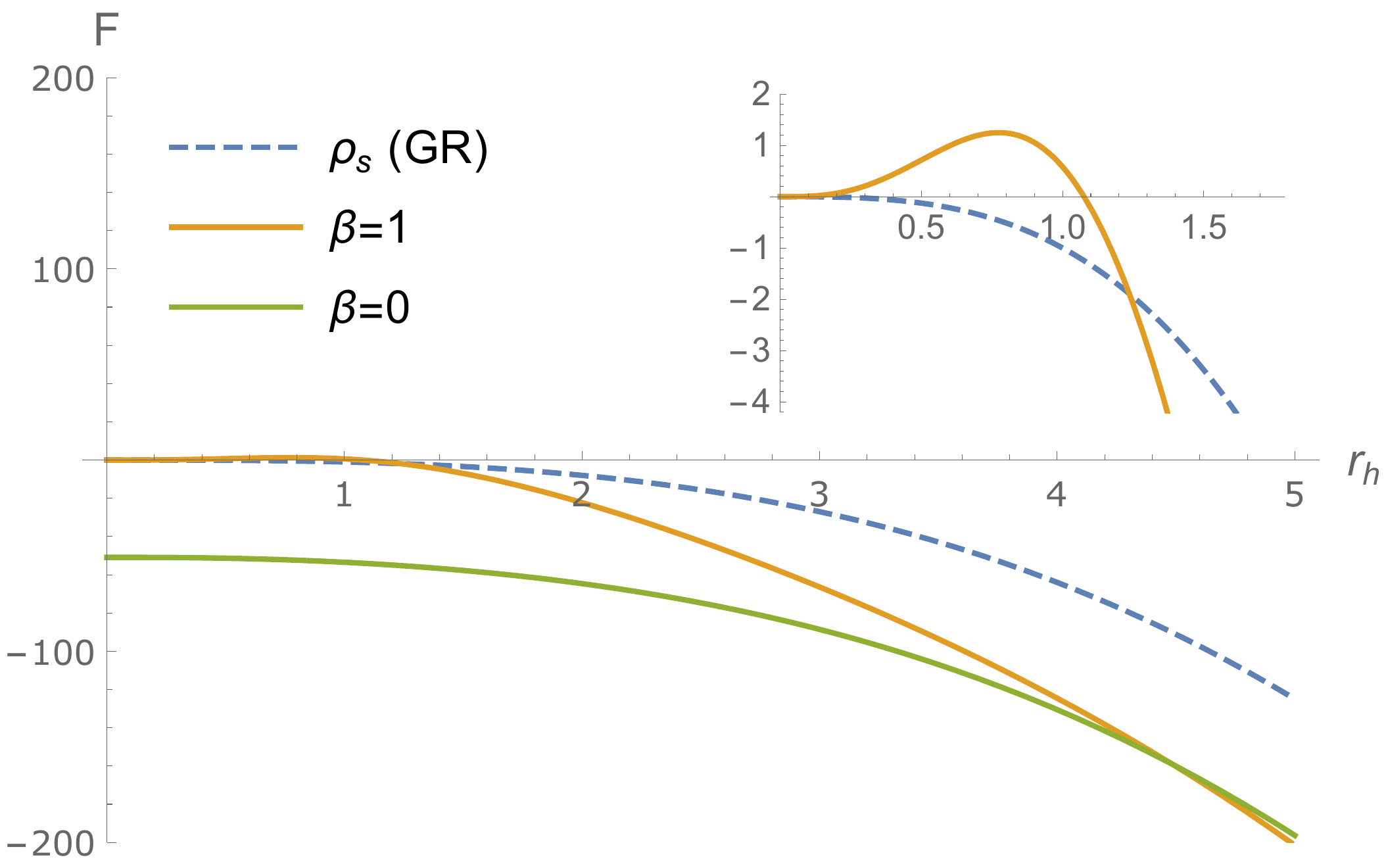}
                \end{minipage}
   \caption{Heat capacity (left panel) and free energy (right panel), per unit length, as functions of the event horizon radii of the black strings. A detailed zoom-in on the region around $r_h \approx 1.07$ for the free energy is presented (right panel)}. The parameter settings are $\ell=0.5$, $\rho_{s}=1.2$, $R_{s}=1.5$, and $\kappa=8\pi$ in Planck units.
    \label{figurasminipg1}
\end{figure}


 Now, we analyze the energy conditions for the dark matter supporting the black string, namely, the Weak Energy Conditions (WEC, $\rho\geq 0$, $\rho+p_i\geq 0$, and, consequently, the Null ones, namely, NEC, for which $\rho+p_i\geq 0$). The index $i$ is for the spatial coordinates, $(1,2,3)\equiv (r,\phi,z)$. The notation $p_r$ is for the radial pressure and $p_{\phi},p_{z}$ are the lateral pressures. The pressures can be obtained by replacing the solution (\ref{metrics}) in the EE (\ref{EME}). As usual, the symmetry of the solutions implies $p_r=-\rho$  and $p_L=p_{\phi}=p_{z}$. Therefore, in order to obtain the lateral pressure, we just need of
 \begin{equation}
\kappa p_{\phi}=G_{\phi}^{\phi}+\Lambda
 \end{equation}
 We finally get
\begin{eqnarray}
p_{\phi}&=&p_z=p_L=\frac{R_s^3 \rho _s \left(r-R_s\right)}{2 r \left(r+R_s\right){}^3}, \\
p_{\phi}&=&p_z=p_L=\frac{4R_s^4 \rho _s \left(r^2-R_s^2\right)}{\left(r^2+R_s^2\right){}^3},
\end{eqnarray}

 for $\beta=1$ and $\beta=0$, respectively. Fig. \ref{figurasminipg2} tells us that both the black string solutions satisfy WEC for all $r$. In other words, dark matter does not behave as an exotic matter.

  \begin{figure}[!ht]
    \centering
    \begin{minipage}{0.45\linewidth}
        \centering
        \includegraphics[width=1.\textwidth]{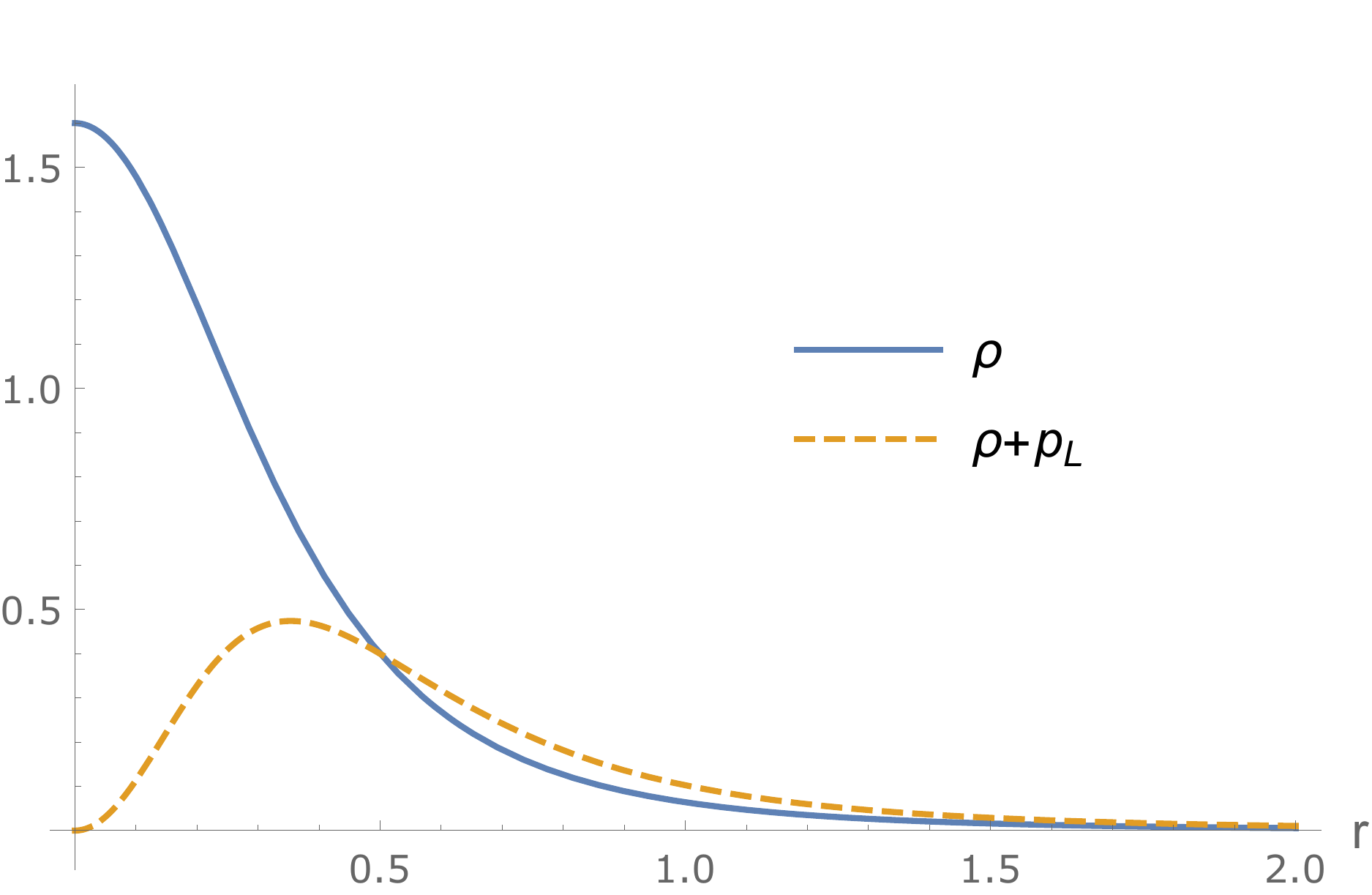}
          \end{minipage}\hspace{0.8cm}
    \begin{minipage}{0.45\linewidth}
        \centering
        \includegraphics[width=1.\textwidth]{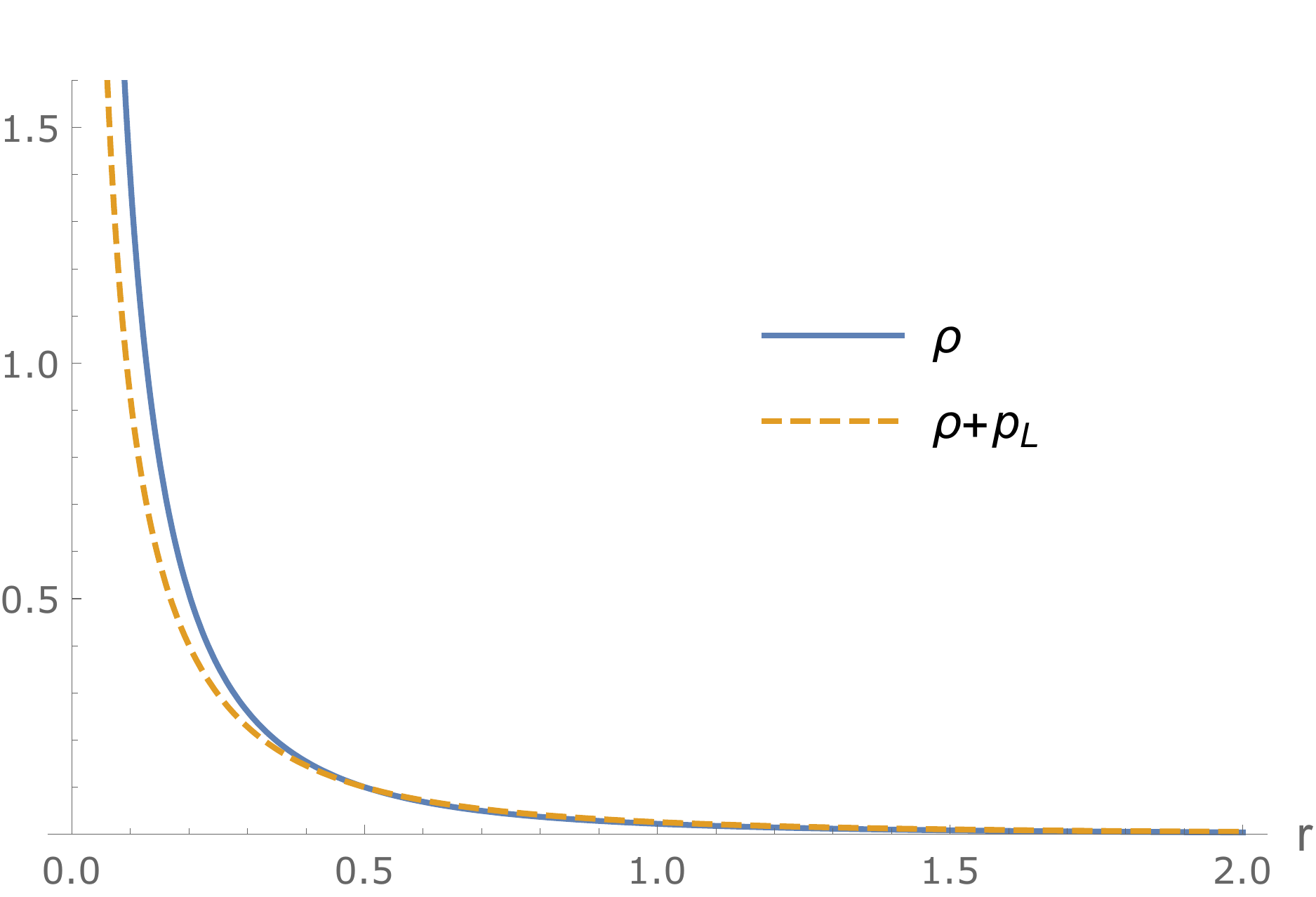}
                 \end{minipage}
   \caption{Energy density, pressures, and their combinations, for $\beta=0$ (left panel) and $\beta=1$ (right panel), as functions of the radial coordinate. The parameter settings are $\rho_{s}=0.4$ and $R_{s}=0.5$, in Planck units.}
    \label{figurasminipg2}
\end{figure}


 \section{Stationary solutions}

We now conceive the stationary counterpart of the black string solutions we have obtained. In order to do this, let us introduce the angular momentum to the static solutions through the following transformation \cite{Lemos2}
\begin{eqnarray}
dt \to \lambda dt-\omega\ell^2 d\phi, \quad d\phi \to \lambda d\phi-\omega dt,\label{transf1}
\end{eqnarray}
where $\lambda$ and $\omega$ are constants to adjust. This prescription yields
\begin{equation}\label{stat1}
ds^2=-[f(r)\lambda^2-r^2\omega^2]dt^2-2[r^2-\ell^2f(r)]\lambda\omega d\phi dt-\frac{dr^2}{f(r)}+[r^2\lambda^2-f(r)\omega^2\ell^4]d\phi^2+\frac{r^2}{\ell^2}dz^2.
\end{equation}
The identification of the black string mass $M$ and angular momentum $J$ can be made through the asymptotic approximation of the static black string metrics given by Eqs. (\ref{metrics}). Hence, following \cite{Lemos2} we obtain
\begin{equation}
    \lambda^2=\frac{M+ \Omega}{-M+ 3\Omega},\; \omega^2\ell^2= \frac{2\left(M- \Omega\right)}{-M+ 3\Omega},
\end{equation}
which implies
\begin{equation}
\mu=\frac{1}{2}\left(-M+ 3\Omega\right).
\end{equation}
Where $\Omega=\sqrt{M^2-\frac{8 J^2}{9\ell^2}}$. Therefore, in the above procedure,  $\mu$ was a parameter that was fixed such that we recover the static solution at infinity. The same must be done with the other parameters of the model. For example, in the charged case it is necessary to identify $\lambda^2 Q^2\to Q^2$\cite{Lemos:1995cm}. The same must be done here, and  we must fix  $\lambda^2\rho_s \to \rho_s $. With this, the final form of the metric is obtained
\begin{eqnarray}
    ds^2&=&-\left[-\frac{2\ell(M+\Omega)}{r}+\frac{r^2}{\ell^2}+\rho_s g(r)\right]dt^2-\frac{8J}{3}\left(\frac{2\ell}{r}-\frac{\rho_{s}}{M+\Omega}g(r)\right)dt d\phi\\
    &&+\frac{dr^2}{\frac{r^{2}}{\ell^{2}}-\frac{2\ell}{r}\left(-M+3\Omega\right)+\rho_{s}\frac{-M+3\Omega}{M+\Omega}g(r)}\\
    &&+\left[r^{2}+\frac{4\ell^{3}(M-\Omega)}{r}-\frac{2(M-\Omega)}{M+\Omega}\rho_{s}\ell^{2}g(r)\right]d\phi^2+\frac{r^2}{\ell^2}dz^2.
\end{eqnarray}
In the above equation, we have
\begin{equation}\label{gdefinition}
g(r)=\left\{ \begin{array}{cc}
\frac{2\kappa   R_s^2}{ 1+(r/R_s)^2}-\frac{2\kappa  R_s^3 \tan ^{-1}\left(\frac{r}{R_s}\right)}{ r}, & \ \ \beta=0;\\
-\frac{\kappa  R_s^3}{r(1+r/R_s)}- \frac{\kappa R_s^3 \log (1+\frac{r}{R_s})}{r}, & \ \ \beta=1. 
\end{array}\right.
\end{equation}
We point out that the above solution recovers the previous one if we choose $J=0$. We also obtain Lemos' vacuum rotating solution if we choose $\rho_s=0$\cite{Lemos2}.
\section{Conclusions}

In this work, we obtained black string solutions sourced by DM with the axisymmetric and isothermal distribution of velocities, in a 4-D spacetime. These solutions depend on a density and a typical reach of DM, tending asymptotically to the usual static and uncharged black string vacuum solution of GR. We have shown that the obtained solutions present only one horizon, similar to what happens in the vacuum solution. Besides this, our analysis showed that a greater concentration of DM increases the size of the horizons. We have also shown that the presence of DM leaves the black strings singular at their axis, which does not occur with the vacuum solution.

From the computation of the event horizons, we have calculated the Hawking temperature of each black string solution and found a linear density of mass (or tension) remnant associated with a vanishing temperature. We have shown that the isotropic black string solution will not have such a remnant if the DM density parameter is less than a critical value, depending on the cosmological constant. On the other hand, the radial solution always will permit the formation of remnants, for any DM density.

We have calculated other thermodynamic quantities associated with the black strings sourced by DM, as the entropy per unit length which presents the same dependence on the event horizon as the vacuum solution. We have also obtained the heat capacity and free energy per unit length, with the black strings solutions exhibiting thermal phase transitions in the presence of DM, as we can see in Fig. \ref{figurasminipg1}, different from what occurs in the vacuum solution. Thus, the heat capacity as a function of the horizon radius reveals regions with local thermal stability and others with instability. On the other hand, the free energy associated with the obtained solutions shows both global thermal stability and instability. We showed that all these quantities (including Hawking temperature) tend asymptotically to those calculated for the vacuum solution of GR. We also analyzed weak (and null) energy conditions from the density and pressures associated with the dark matter, which are finite at the origin only in the isotropic case, and concluded that it does not behave like an exotic fluid, for both the considered solutions.

Finally, we have obtained the corresponding stationary solutions, and from the asymptotic behavior of the static counterparts, we have determined the expressions for both the mass and angular momentum of the black strings, showing that the tension $\mu$ of the obtained solutions depends on these parameters. By vanishing the angular momentum, we re-obtain the static solutions previously studied, and by eliminating the DM parameters, we retrieve Lemos' stationary black string solution, as expected. 





\acknowledgments{The authors G. A., C. R. M, V. B. B., and M. S. C. would like to thank Conselho Nacional de Desenvolvimento Cient\'{i}fico e Tecnol\'{o}gico (CNPq) and Fundação Cearense de Apoio ao Desenvolvimento Científico e
Tecnológico (FUNCAP) through PRONEM PNE0112- 00085.01.00/16, for the partial financial support. H.S.V. was funded by the Alexander von Humboldt-Stiftung/Foundation (Grant No. 1209836). This study was financed in part by the Coordena\c c\~{a}o de Aperfei\c coamento de Pessoal de N\'{i}vel Superior - Brasil (CAPES) - Finance Code 001. This study was financed in part by the Conselho Nacional de Desenvolvimento Cient\'{i}fico e Tecnol\'{o}gico -- Brasil (CNPq) -- Research Project No. 150410/2022-0.
}



\begin{thebibliography}{99}

\bibitem{Levi} Levi-Civita, Rend. Acc. Lincei, 26, 317 (1917).
\bibitem{Weyl} H. Weyl, Annalen Phys., 54, 117 (1917).
\bibitem{Chazy} J. Chazy, Bull. Soc. Math. France, 52, 17 (1924).
\bibitem{Curzon} H. E. J. Curzon, Proc. R. Soc. London, 23, 477 (1924).
\bibitem{Lewis} T. Lewis, Proc. R. Soc. London, A 136, 176 (1932).
\bibitem{Vilenkin}  A. Vilenkin, Phys. Rep, 121, 263 (1985).
\bibitem{Gott} J. R. Gott III, Phys. Rev. Lett. {\bf66}, 1126 (1991).
\bibitem{Godel} K. G\"{o}del, Rev. Mod. Phys. 21, 447 (1949).
\bibitem{Krasinski} A. Krasinski, Acta Cosmologica, Zesz. 7, p. 133 (1978).
\bibitem{duff1988} M. J. Duff, T. Inami, C. N. Pope, E. Sezgin, and K. S. Stelle, Nucl. Phys. {\bf B297}, 515 (1988).
\bibitem{emparanPRL}R. Emparan, H. S. Reall, Phys. Rev. Lett. {\bf88}, 101101 (2002).
\bibitem{ijmpa2011}S. Bellucci, B. N. Tiwari, Int. J.  Modern Physics {\bf A26}, 5403 (2011).
\bibitem{Thorne} K.S. Thorne, in Magic without magic, ed. J. R. Klander (Freeman, San Francisco, 1972).
\bibitem{Lemos1} J.P.S. Lemos, Class. Quantum Gravity {\bf12}, 1081 (1995).
\bibitem{Lemos2} J. P. S. Lemos, Phys.Lett. {\bf B353}, 46 (1995).
\bibitem{Ghosh} L. Tannukij, P. Wongjun, and S. G. Ghosh, Eur. Phys. J. C {\bf77}, 846 (2017).
\bibitem{Hendi} S. H. Hendi, H. Zarei, M. Faizal, B. Pourhassan, and Z. Armanfard, Nucl. Phys. {\bf B965}, 115362 (2021).
\bibitem{Singh} D. V. Singh, M. S. Ali, and S. G. Ghosh, Int.J. Mod. Phys. {\bf D27}, 12, 1850108 (2018).
\bibitem{Pino} A. Cisterna, C. Corral, and S. Pino, Eur. Phys. J. {\bf C79}, 400(2019).
\bibitem{Ahmad} A. Sheykhi, J. High E.Phys. {\bf2020}, 31 (2020).
\bibitem{Sabir} M. S. Ali, F. Ahmed, and S. G. Ghosh, Ann. of Phys. {\bf412}, 168024 (2020).
\bibitem{Celio} C. R. Muniz, H. R. Christiansen, M. S. Cunha, and J. Furtado, Ann. of Phys. {\bf443}, 168980 (2022).
\bibitem{Planck:2018vyg} N.~Aghanim {\it et al.}, Astron. Astrophys. {\bf641}, A6 (2020).
\bibitem{Trujillo-Gomez:2010jbn} S.~Trujillo-Gomez, A.~Klypin, J.~Primack, and A.~J.~Romanowsky, Astrophys. J. {\bf742}, 16 (2011).
\bibitem{Kim} S. E. Hong, D. J., H. S. Hwang, and J. Kim, The Astrophys. J. 913 (1) 76 (2021).
\bibitem{Callum} C. T. Donnan, R. Tojeiro, and K. Kraljic, Nat Astron (2022), doi.org/10.1038/s41550-022-01619-w.
\bibitem{Eisenstein} D. J. Eisenstein, A. Loeb, and E. L. Turner, The Astroph. J. {\bf475}, 421 (1997).
\bibitem{Rodrigo} R. Calderón, R. Gannouji, B. L'Huillier, and D. Polarski, Phys. Rev. {\bf D 103}, 023526 (2021).
\bibitem{Lemos:1995cm}
J.~P.~S.~Lemos and V.~T.~Zanchin,
Phys. Rev. D \textbf{54}, 3840-3853 (1996)
doi:10.1103/PhysRevD.54.3840
[arXiv:hep-th/9511188 [hep-th]].
	
\bibitem{Fatima:2011dr}
A.~Fatima and K.~Saifullah,
Astrophys. Space Sci. \textbf{341}, 437-443 (2012)
doi:10.1007/s10509-012-1098-2
[arXiv:1108.1622 [gr-qc]].

 \end{thebibliography}
 \end{document}